\documentclass{egpubl}
\usepackage{eurova2025}

\ConferenceSubmission   
\WsPaper           

\usepackage[T1]{fontenc}
\usepackage{dfadobe}  

\usepackage{cite}  
\BibtexOrBiblatex
\electronicVersion
\PrintedOrElectronic
\ifpdf \usepackage[pdftex]{graphicx} \pdfcompresslevel=9
\else \usepackage[dvips]{graphicx} \fi

\usepackage{egweblnk} 
\usepackage{tikz}
\usepackage{amssymb}
\usepackage{multirow}
\usepackage{array}
\usepackage{rotating}
\usepackage{makecell}
\usepackage{booktabs}

\usepackage{color}
\definecolor{seagreen}{rgb}{0,0.65,0.5}
\definecolor{RED}{rgb}{1,0,0}
\definecolor{BLUE}{rgb}{0,0,1}
\definecolor{GREEN}{rgb}{0,0.8,0}
\definecolor{michaelcolor}{rgb}{0,0.65,0.5}
\definecolor{MAGENTA}{rgb}{0,0.7,0.7}
\definecolor{iScore}{HTML}{F6AAFF}
\definecolor{Comparator}{HTML}{FFF08A}

\definecolor{dSpaceExplanation}{rgb}{0.631, 0.769, 0.957} 
\definecolor{dSpaceGroundTruth}{rgb}{0.702, 0.839, 0.655} 
\definecolor{dSpaceLLMValue}{rgb}{1.000, 0.804, 0.529}

\newcommand\changed[1]{{\color{black}#1}}

\tikzset{bubble/.style={
    draw=black, 
    rounded corners=0pt, 
    inner sep=1pt, 
    minimum height=10pt,
    text height=1.2ex, 
    text depth=0.1ex
}}
\newcommand{\singleItem}{%
    \tikz[baseline=(X.base)] 
        \node[bubble, fill=gray!30] 
        (X) {1 Item};%
}
\newcommand{\multipleItem}{%
    \tikz[baseline=(X.base)] 
        \node[bubble, fill=gray!30] 
        (X) {m Items};%
}
\newcommand{\oneAttribute}{%
    \tikz[baseline=(X.base)] 
        \node[bubble, fill=gray!30] 
        (X) {$1 \times 1$ Attribute};%
}
\newcommand{\multipleAttribute}{%
    \tikz[baseline=(X.base)] 
        \node[bubble, fill=gray!30] 
        (X) {$1 \times m$ Attributes};%
}
\newcommand{\allAttribute}{%
    \tikz[baseline=(X.base)] 
        \node[bubble, fill=gray!30] 
        (X) {$m \times m$ Attributes};%
}

\newcommand{\val}{%
    \begin{tikzpicture}[baseline={([yshift=-0.5ex]current bounding box.center)}]
        \draw[fill=dSpaceLLMValue] (0,0) rectangle (1em,1em);
        \node[font=\bfseries] at (0.5em,0.5em) {V};
    \end{tikzpicture}%
}
\newcommand{\explanation}{%
    \begin{tikzpicture}[baseline={([yshift=-0.5ex]current bounding box.center)}]
        \draw[fill=dSpaceExplanation] (0,0) rectangle (1em,1em);
        \node[font=\bfseries] at (0.5em,0.5em) {E};
    \end{tikzpicture}%
}
\newcommand{\valg}{%
    \begin{tikzpicture}[baseline={([yshift=-0.5ex]current bounding box.center)}]
        \draw[fill=dSpaceGroundTruth] (0,0) rectangle (1em,1em);
        \node[font=\bfseries] at (0.5em,0.5em) {$V_{g}$};
    \end{tikzpicture}%
}

\newcommand{\valExample}[2]{%
    \tikz[baseline=(X.base)] 
        \node[fill=#1, rounded corners=5pt, inner sep=1pt] 
        (X) {#2};%
}

\title[Design Space for Critical Validation of LLM Tabular Data]%
      {A Design Space for the Critical Validation of LLM-Generated Tabular Data}

\author[M. Sachdeva, \& C. Narayanan, \& M. Wiedenkeller ,\& J. Sedlakova, \& J. Bernard]
{\parbox{\textwidth}{\centering Madhav Sachdeva$^{1}$\orcid{0000-0002-9840-7735},
        Christopher Narayanan$^{1}$\orcid{0009-0009-8500-9656},
        Marvin Wiedenkeller$^{1}$\orcid{0009-0002-1853-3457},
        Jana Sedlakova$^{1,2}$\orcid{0000-0002-6887-5941},
        and Jürgen Bernard$^{1,2}$\orcid{0000-0001-8741-9709} 
        }
        \\
{\parbox{\textwidth}{\centering $^1$University of Zurich, Switzerland; $^2$Digital Society Initiative, Zürich, Switzerland \\
       }
}
}

\begin{document}

\teaser{
\vspace{-5mm}
 \includegraphics[width=0.85\linewidth]{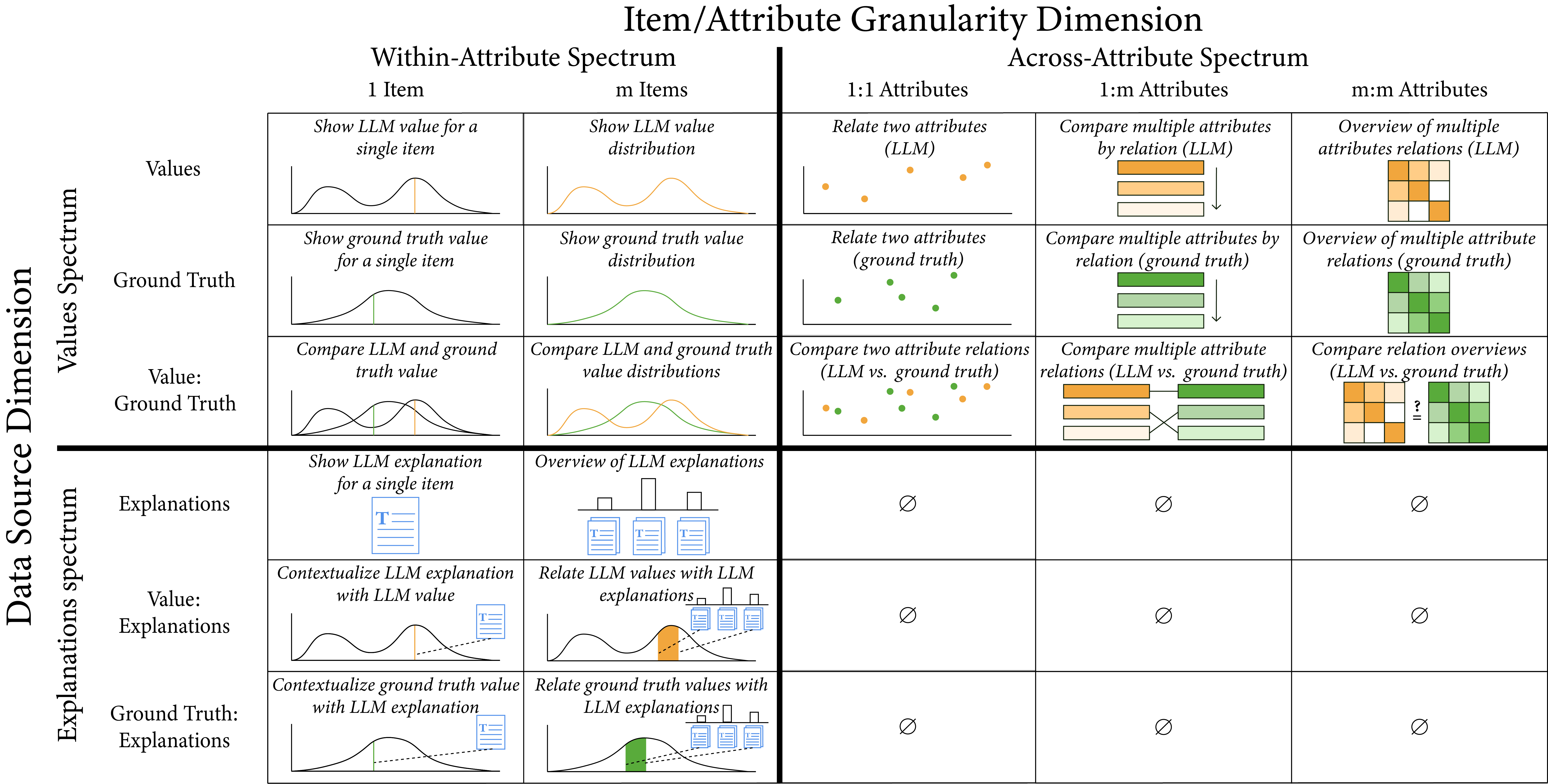}
 \centering
  \caption{Design space for the critical validation of LLM-generated tabular data. The x-dimension describes the analysis granularity, ranging from \changed{within attributes to across attributes}. The y-dimension structures \changed{approaches by the data used for validation, including generated values}, explanations, ground truth data, and \changed{their} combinations. Visual representations per cell exemplify numerical data use, while the design space is generalizable for mixed data. \changed{Lower right: validations across attributes including explanations are so far uncharted.}}
\label{fig:teaser}
}

\maketitle
\begin{abstract}
   LLM-generated \changed{tabular data is creating new opportunities for data-driven applications} in academia, business, and society.
   \changed{To leverage benefits like missing value imputation, labeling, and enrichment with context-aware attributes, LLM-generated data needs a critical validation process.
   The number of pioneering approaches is increasing fast, opening a promising validation space that, so far, remains unstructured.}
   We present a design space for the critical validation of LLM-generated tabular data with two dimensions:
   First, the Analysis Granularity dimension--from within-attribute (single-item and multi-item) to across-attribute perspectives ($1 \times 1$, $1 \times m$, and $n \times n$).
   Second, the Data Source dimension--differentiating between LLM-generated values, ground truth values, explanations, and their combinations. We \changed{discuss} analysis tasks for each dimension cross-cut, map \changed{19} existing validation approaches, and discuss the characteristics of two approaches in detail, demonstrating descriptive power.   


\printccsdesc  
\end{abstract}  


\section{Introduction}

Large language models (LLMs) have evolved from simple text prediction to generating context-aware output across diverse data types, \changed{with data-generation qualities comparable to state-of-the-art deep learning methods~\cite{miletic2024assessing}.}
\changed{LLMs are increasingly} used to enrich tabular datasets with contextually relevant values for items and attributes, and create new items relevant to analytical needs.
Common LLM-based data enrichment methods include imputing missing values, labeling data items, and adding semantically rich attributes~\cite{fang2024large}.
\changed{However, critical validation is essential to ensure the usability, plausibility, and reliability of LLM-generated tabular data.}
Key problem areas \changed{in LLM-generated data} include subjective quality~\cite{kim2024evallm}, lack of reproducibility\changed{~\cite{coscia2024iscore}}, \changed{factuality hallucinations (e.g., inconsistency with real-world facts), faithfulness hallucinations (e.g., deviating from intended prompt), epistemic opacity (black-box nature), biases, and adoption of human misconceptions~\cite{malin2024review}}. 
\changed{Further challenges include inconsistencies in attribute relationships appearing as logical contradictions, or inconsistencies in logical reasoning steps~\cite{liu2024aligning}. 
Research also identified LLM biases in choice preferences (e.g., favoring first position, favoring style of input data  (e.g., preferring longer responses), and egocentric bias (e.g., favoring their own responses)~\cite{li2023prd}.}
Without critical validation, integrating such data into analytical workflows risks propagating errors, which misinforms decision-making processes\changed{\cite{park2024assessing}}.

\changed{Validation methods differ in their degree of human and computational involvement, creating opportunities for visual analytics (VA) research.
Fully automated methods, like statistical measures \cite{sun2024comprehensive} and self-validating techniques like Retrieval-Augmented Generation (RAG) \cite{asai2023self}, offer scalability but reduce transparency.
Mixed-initiative methods, such as LLM-as-a-Judge~\cite{zheng2023judging} and human-calibrated evaluations \cite{shankar2024validates,kahng2024llm}, combine automation with human feedback. 
Direct human evaluation maximizes interpretability, but is resource-intensive and less scalable \cite{chiang-lee-2023-large}.
}

\changed{While validation is well-established for unstructured textual data~\cite{long2024llms}, recent work increasingly targets tabular data, showing high heterogeneity in methods.
Validation strategies include the statistical alignment with benchmark data \cite{sun2024comprehensive}, LLM-as-a-Judge \cite{zheng2023judging}, explanation-based methods \cite{huang2023can}, and ground truth comparisons \cite{wang2024human}.
Ground truth may be human-labeled or from external datasets \cite{wang2024human}. Human-labeled data is often used in mixed-initiative approaches \cite{kim2024evallm, kahng2024llm, shankar2024validates}.
Approaches operate at different analysis granularities, trading accuracy for scalability, ranging from fine-grained single-item assessments \cite{kim2024evallm}, multi-item or entire attributes for aggregated metrics \cite{cabrera2023zeno, coscia2023knowledgevis}, to coarse assessments across attributes \cite{tenney2020language}.
Some provide textual explanations for single items \cite{huang2023can}, or multiple items to show systematic biases \cite{kahng2024llm}.
Approaches also differ across workflow phases: LLM-based data generation (prompt engineering)~\cite{miletic2024assessing}, core validation (output assessments)~\cite{kahng2024llm}, and evaluating LLM outputs in domain-specific downstream applications~\cite{coscia2023knowledgevis}.
}
Contextual factors such as the LLM model, application domain, and user group introduce further variations.

\changed{To understand this emerging and multifaceted validation space, several challenges must be addressed. 
Existing approaches operate in an unstructured space without an overarching perspective, leaving key characteristics implicit.
No comprehensive meta-analysis compares validation methods, leaving differences and commonalities unexplored.
The lack of structured analysis prevents easy identification of gaps and the guided selection of appropriate strategies.
A unified framework is needed to organize methods, identify gaps, and guide future research.
}

We introduce a design space that systematically structures validation approaches for LLM-generated tabular \changed{mixed} data.
Derived from a systematic literature review, it is defined along two key dimensions: 
The \textit{Item and Attribute Granularity Dimension} captures the validation granularity, within and across attributes.
The \textit{Data Source Dimension} \changed{covers} core validation data sources: LLM-generated values, ground truth, LLM-explanations, and their combinations. 
At the intersection of these dimensions, we identify essential validation tasks that guide validation methods for LLM-generated tabular data. 
Our contributions are:
\begin{enumerate}
    \item \textbf{A design space} for the critical validation of LLM-generated structured tabular data. 
    It structures and describes methods by their commonalities and differences, and guides designers and developers toward novel validation strategies.
    \item \textbf{A mapping of existing validation approaches} onto this design space, demonstrating its descriptive power. It offers a systematic characterization of approaches and the identification of gaps.
    \item \textbf{A discussion on key insights} based on critical reflection, including identified patterns, uncharted subspaces, and additional structural characteristics to inform future research. 
\end{enumerate}

\changed{Our design space maps the validation landscape to support research into robust, scalable, and interpretable validation methods.
It aims to enhance the plausibility, reliability, and utility of LLM-generated tabular data in data analysis and decision-making workflows, accounting for different types of human and algorithmic involvement.
It supports researchers, system designers, and tool developers to build VA interfaces for critical data validation.
}

\section{Design Space} 

\subsection{Methodology \changed{and Overview}}
\label{sec:methodology}
\changed{We conducted a} systematic literature search, analysed the identified related work, and discussed it among the authors. \changed{Three authors identified and categorized validation approaches for LLM-generated tabular data via Google Scholar search with the terms ‘LLM tabular data validation’, ‘LLM human-in-the-loop validation’, ‘Explanation LLM-generated tabular data’, ‘Explainable LLM interactive visualizations’ and ‘LLM scoring’, followed by a forward and backward search, with a focus on the papers EvalLM~\cite{kim2024evallm}, LLM Comparator~\cite{kahng2024llm}, and two surveys by Brasoveanu et al.~\cite{brasoveanu2024visualizing} and Long et al.~\cite{long2024llms}.}
We excluded works published before 2017 (prior to transformer-based LLM architectures), those not written in English, \changed{mentioning validation only superficially}, or addressing only textual data \changed{(like EvalGen~\cite{shankar2024validates})}.
Borderline cases were discussed to reach consensus. 
In total, we identified \changed{19 relevant works as the basis for design space ideation; we refer to the supplemental material for \changed{their characterization}. 
The main visualizations used by existing approaches are bar charts (6), followed by scatter plots (3) and heat maps (3).}

\changed{We aimed to support the systematic study of existing approaches and the identify gaps and opportunities, in both numerical and categorical tabular data.
We chose a 2D structure for spatialization, using expressive, independent dimensions that reveal validation tasks supported by VA.
Both dimensions are discrete to allow systematic cross-cuts and clear mapping of approaches in a table.
}

Our \changed{design space} structures validation methods for LLM-generated tabular data by two key dimensions: \textit{Data Source}, and \textit{Item and Attribute Granularity}, reflecting \textit{what} and \textit{why} experts analyze~\cite{munzner2014visualization}. 
The \textit{Data Source} dimension spans \changed{data types serving as inputs essential for validation reasoning, consisting of (combinations of)} LLM-generated data, explanations, and (human) ground truth.
The \textit{Item and Attribute Granularity} dimension \changed{distinguishes approaches by analysis granularity, from fine-grained item-based validations and coarse attribute-based assessments, offering different trade-offs between accuracy and scalability}.
Figure \ref{fig:teaser} illustrates the design space with important analysis tasks for cross-cutting dimensions.
\changed{Each task reflects the need for specific visual reasoning, depending on Data Source and analysis granularity. 
For most tasks, existing approaches already indicate the need for interactive human judgment enabled through VA support.}
\changed{We exemplify the design space with visual identifiers for numerical data due to its intuitiveness, while the space is mixed-data capable.}
Table \ref{tab:design-space} displays the mapping of existing approaches onto the design space. 


\subsection{Design Space Dimension: Data Source}

The Data Source dimension refers to the input data used for validation tasks.
We use visual encodings for the three types of Data Sources: LLM-generated values (\val), ground truth values (\valg) and LLM-generated explanations (\explanation), and their combinations.

\begin{itemize}
    \item \textbf{Data Value \val{}}: is the most granular information unit requested from the LLM. Data values belong to an item and have an attribute type defined by its associated attribute schema. 
     \item \textbf{Ground truth Value \valg{}}: represents a data value for an item that is known or assumed to be true. This can be provided by pre-existing data or elicited from human externalized knowledge. 
     \item \textbf{Explanations \explanation{}}: are textual justifications provided by an LLM about its generated data values. \changed{This explanation spectrum enables to contextualize data values or assess their semantic meaning with VA relation-discovery, per item or more condensed}.
\end{itemize}

\changed{
A key observation from many approaches highlight a key distinction: value-based validation (\val{}, \valg{}) relies on measurable data, statistical reasoning, and alignment checks, while explanation-based (\explanation{}) validation requires subjective judgments of plausibility, consistency, and bias.
This distinction directly impacts the types of visualization tasks and the design of VA solutions.
}

\begin{table*}[ht]
\small
\centering
\renewcommand{\arraystretch}{1.0} 
\resizebox{0.90\textwidth}{!}{%
\begin{tabular}{|p{0.1cm}|>{\raggedright\arraybackslash}p{1.6cm}|>{\raggedright\arraybackslash}p{4.3cm} |>{\raggedright\arraybackslash}p{3.9cm}!{\vrule width 1.5pt}>{\raggedright\arraybackslash}p{2.8cm}|>{\raggedright\arraybackslash}p{2.8cm}|>{\raggedright\arraybackslash}p{2.5cm}|>{\raggedright\arraybackslash}p{2.5cm}|}
\hline
\multicolumn{2}{|c|}{} & \multicolumn{5}{c|}{\textbf{Item/Attribute Granularity Dimension}} \\
\hline
\multicolumn{2}{|c|}{} &
\textbf{\singleItem{}} &
\textbf{\multipleItem{}} &
\textbf{\oneAttribute{}} &
\textbf{\multipleAttribute{}} &
\textbf{\allAttribute{}} \\
\Xhline{1pt}
\multirow{6}{*}{\rotatebox{90}{\textbf{Data Source Dimension}}} & \textbf{Values \val{}} & Show LLM value for a single item~\cite{kahng2024llm, kim2024evallm, cheng2024interactive, tenney2020language, coscia2023knowledgevis} & Show LLM value distribution~\cite{kahng2024llm, pan2024human, kim2024evallm, cabrera2023zeno, cheng2024interactive, tenney2020language, coscia2023knowledgevis} & Relate two attributes (LLM)~\cite{kahng2024llm, kim2024evallm, tenney2020language} & Compare multiple attributes by relation (LLM) & Overview of multiple attribute relations (LLM)\cite{tenney2020language}\\
\cline{2-7}
& \textbf{Ground Truth \valg{}} & Show ground truth value for a single item~\cite{ pan2024human, tenney2020language} & Show ground truth value distribution~\cite{coscia2024iscore, tenney2020language} & Relate two attributes (ground truth)~\cite{coscia2024iscore, tenney2020language}  & Compare multiple attributes by relation (ground truth) & Overview of multiple attribute relations \cite{tenney2020language} \\
\cline{2-7}
 & \textbf{Value : Ground Truth \val{}:\valg{}} & Compare LLM and ground truth value~\cite{pan2024human, strobelt2022interactive, mishra2025promptaid, cabrera2023zeno, zhu2023can, hegselmann2023tabllm, fan2024finding, shankar2024spade, geng2025generating, tenney2020language, coscia2024iscore} & Compare LLM and ground truth value distributions \cite{coscia2024iscore, pan2024human,  strobelt2022interactive, mishra2025promptaid, cabrera2023zeno, zhuang2023beyond, borisov2022language, tenney2020language} &\changed{Compare two attribute relations (LLM vs. ground truth)}~\cite{fan2024finding, borisov2022language, tenney2020language} &  Compare \changed{multiple attribute relations} (LLM vs. ground truth)~\cite{fan2024finding, borisov2022language} & Compare relation overviews (LLM vs. ground truth) \\
\Xcline{2-7}{1.5pt}
 & \textbf{Explanations \explanation{}} & Show LLM explanation for a single item & Overview of LLM explanations & & & \\
\cline{2-7}
 & \textbf{Value : Explanations \val{}:\explanation{}} & Contextualize LLM explanation with LLM value~\cite{kahng2024llm, coscia2024iscore, pan2024human, kim2024evallm, strobelt2022interactive, mishra2025promptaid, cheng2024interactive, becker2024cycles, schroeder2025large, tenney2020language} & Relate LLM values with LLM
explanations~\cite{kahng2024llm, becker2024cycles} & & & \\
\cline{2-7}
 & \textbf{Ground Tr. : Explanations \valg{}:\explanation{}} & Contextualize ground truth value with LLM explanation~\cite{becker2024cycles, tenney2020language} & Relate ground truth values with
LLM explanations~\cite{becker2024cycles}  & & & \\
\hline
\end{tabular}
}
\caption{Design space tasks across Item and Attribute Granularity and Data Source Dimension. While the design space accommodates values across all item-attribute granularities, explanations are primarily effective when applied to an item granularity.}
\label{tab:design-space}
\vspace{-3mm}
\end{table*}

\subsection{Design Space Dimension: Item and Attribute Granularity}
This dimension characterizes the validation granularity,
in a spectrum from fine-grained \textit{Item-based} validations within an attribute, to more coarse \textit{Attribute-based} validation, across attributes.
\begin{itemize}
    \item \singleItem{}: the most atomic granularity of validation is for a single item \changed{(and attribute), preferring accuracy over scalability}.
    \item \multipleItem{}: multiple items are taken into account for a single attribute. This includes the assessment of summaries and aggregates, and the comparative analysis between item values.
    \item \oneAttribute{}: raising the granularity to an across-attributes perspective, one class of validation approaches is on relation discovery tasks between an attribute pair.
    \item \multipleAttribute{}: for some validation tasks, users take multiple attributes into account, e.g., to assess the strengths of relations of attributes to a focus attribute.
    \item \allAttribute{}: the coarsest granularity is the brute-force assessment of all relations between all attributes \changed{at scale}.
\end{itemize}

\subsection{Validation Tasks at the Crosscut of Dimensions} 
The systematic combination of the \textit{Data Source} and the \textit{Item and Attribute Granularity} dimension opens the perspective of discrete cells of the design space.
For each cell, \changed{we present the dominating validation task, according to existing approaches (see Table~\ref{tab:design-space}). 
Blank cells at the lower right indicate non-allocated regions (e.g., aggregating explanations across (multiple) attributes simultaneously.)}
\changed{Figure~\ref{fig:teaser} provides detailed task descriptions and visual exemplifications, to align validation tasks with abstract analysis/visualization tasks, e.g., "Show LLM value for a single item" supports \textit{Lookup} whereas "Compare multiple attributes by relation" supports \textit{Compare}.
For each cross-cut, we include expected analysis outcomes and example analysis questions.}

\subsubsection{\protect\singleItem{} Focus }
\changed{Expected analysis outcome: identify and clarify discrepancies between human-expected and LLM-generated values for a single item}.
\changed{Example analysis question: “Is the LLM-generated happiness index for the city of Luxembourg plausible?”}

\val{} $\times$ \singleItem{}: 
\textbf{Show LLM value:} 
The isolated inspection of a single LLM-generated value enables users to assess its plausibility and relevance.
Single-value assessment is often supported in the context of value distributions, e.g., to enable comparative analysis. 
\\
\valg{} $\times$ \singleItem{}:
\textbf{Show ground truth value:} 
Allows users to create, inspect, or refine a ground truth value, which can be later used for comparative assessment with LLM-generated values.
\\
\val{} $\times$ \valg{} $\times$ \singleItem{}:
\textbf{Compare LLM value and ground truth value:} 
Users can compare the LLM-generated value with the ground truth value, enabling the assessment of value alignment/deviation.
\\
\explanation{} $\times$ \singleItem{}:
\textbf{Show LLM explanation:} 
Users can engage with explanations for a single LLM-generated value. 
This can potentially show flaws in the LLM's reasoning or justification.
\\
\explanation{} $\times$ \val{} $\times$ \singleItem{}:
\textbf{Contextualize LLM explanation with LLM value:} 
Displaying the LLM-generated value and its explanation helps assess its justification and reveal inconsistencies. 
This pairing offers insight into the LLM’s logic, indicating whether it considers relevant factors. 
It also highlights weaknesses when explanations are vague, uncertain, or misaligned with generated values.
\\
\valg{} $\times$ \explanation{} $\times$ \singleItem{}: 
\textbf{Contextualize ground truth value with LLM explanation:} 
This task enables the assessment of the LLM-generated explanation by using the ground truth value as a trusted frame of reference, e.g., to shed light on the reasoning of the LLM.

\subsubsection{\protect\multipleItem{} Focus}
\changed{Expected analysis outcome: identify and clarify discrepancies between human-expected and LLM-generated value distributions for multiple items}.
\changed{Example analysis question: “Are LLM-generated unemployment rates biased toward a specific range?”}

\val{} $\times$ \multipleItem{}:
\textbf{Show LLM value distribution:} 
Displaying the distribution of LLM-generated values, possibly in comparison to the entire item value distribution, can reveal subset characteristics, like potential biases or a lack of diversity.
\\
\valg{} $\times$ \multipleItem{}:
\textbf{Show ground truth value distribution:} 
Show ground truth value distribution to reveal subset patterns and potential shifts from the overall distribution, helping to assess biases. 
\\
\explanation{} $\times$ \multipleItem{}:
\textbf{Overview of LLM explanations:} 
Understanding explanation patterns occurring across multiple items helps to cross-validate explanations, understand rationales that drive LLMs when generating values, and helps to assess the LLM's consistency.
\\
\val{} $\times $\explanation{} $\times$ \multipleItem{}:
\textbf{Relate LLM values with LLM explanations:} 
Relating multiple LLM-generated values with explanations assesses justification consistency, reveals discrepancies across items, and identifies potential pitfalls or biases in the reasoning.
\\
\val{} $\times $\valg{} $\times$ \multipleItem{}:
\textbf{Compare LLM and ground truth value distributions:} 
Assessing the alignment of these two value distributions provides statistical insights, revealing discrepancies, potential biases, or failures in LLM’s understanding of key aspects. 
\\
\valg{} $\times $\explanation{} $\times$ \multipleItem{}:
\textbf{Relate ground truth values with LLM explanations:}
This relation-discovery task allows users to
assess the general alignment between ground truth and LLM explanations for the identification of potential pitfalls or biases in the reasoning.

\subsubsection{\protect\oneAttribute{} Focus}
\changed{Expected analysis outcome: clarity and plausibility of a LLM-generated attribute, in relation to an existing attribute}.
\changed{Example analysis question: “Are happiness scores generated by the LLM positively correlated with my income attribute?”}

\val{} $\times$ \oneAttribute{}:
\textbf{Relate two attributes (LLM-generated):} 
Relation-discovery between two attributes, e.g., a LLM-generated and an existing attribute to show interesting dependencies.
\\
\valg{} $\times$ \oneAttribute{}:
\textbf{Relate two attributes (ground truth):} 
Relation-discovery between two ground truth attributes, e.g., to identify dependencies or redundancies for feature selection.
\\
\valg{} $\times$ \val{} $\times$ \oneAttribute{}:
\textbf{\changed{Compare two attribute relations} (ground truth with LLM-generated):} 
Relate ground truth and LLM-generated attributes to assess alignment and agreement. 
This helps to gain trust in whether LLMs can reproduce what is already known.

\subsubsection{\protect\multipleAttribute{} Focus}
\changed{Expected analysis outcome: clarity about the relation between the LLM-generated attribute and other (existing) attributes}.
\changed{Example analysis question: “Which attributes are most positively/negatively correlated with the LLM-generated happiness index?”}

\val{} $\times$ \multipleAttribute{}:
\textbf{Compare multiple LLM-generated attributes by \changed{relation}:} 
Relation-discovery between LLM-generated attributes can, e.g., put one LLM-generated attribute in focus, allowing for ranking other attributes by their relation strength.
\\
\valg{} $\times$ \multipleAttribute{}:
\textbf{Compare multiple ground truth attributes by \changed{relation}:} 
Relation-discovery between ground truth attributes can, e.g., put one attribute in focus, allowing for ranking other attributes by their relation strength.
\\
\val{} $\times$ \valg{} $\times$ \multipleAttribute{}:
\textbf{\changed{Compare multiple attribute relations }(LLM vs ground truth):}
Users compare the relation-discovery outcomes of LLM-generated attributes with those for ground truth attributes to identify discrepancies and biases.
\subsubsection{\protect\allAttribute{} Focus}
\changed{Expected analysis outcome: clarity on the relations between all LLM-generated attributes and all existing attributes.}
\changed{Example analysis question: “Do relations in LLM-generated data align with the ground truth and do they enrich existing data for downstream use?”}

\val{} $\times$ \allAttribute{}:
\textbf{Overview of multiple LLM-generated attributes and their relations:}
This relation-discovery task applies a brute-force approach to relate all attributes to each other, typically leading to a matrix display of relations.
\\
\valg{} $\times$ \allAttribute{}:
\textbf{Overview of multiple ground truth attributes and their relations:} 
This task provides users with the full overview of all relations between ground truth attributes.
\\
\val{} $\times$ \valg{} $\times$ \allAttribute{}:
\textbf{Compare relation overviews (LLM with ground truth):} 
This task compares all relations of LLM-generated attribute pairs with all relations of ground truth attribute pairs to gain an understanding of their agreement. 



\section{Validation Examples}
We demonstrate the descriptive power of the design space by mapping tasks from two existing approaches \changed{and show how their workflows map into the design space.}
Arrows between tasks illustrate the workflows described in these approaches (see Figure~\ref{fig:validation_example}).

\begin{figure}[t]
\centering\includegraphics[width=1.0\linewidth]{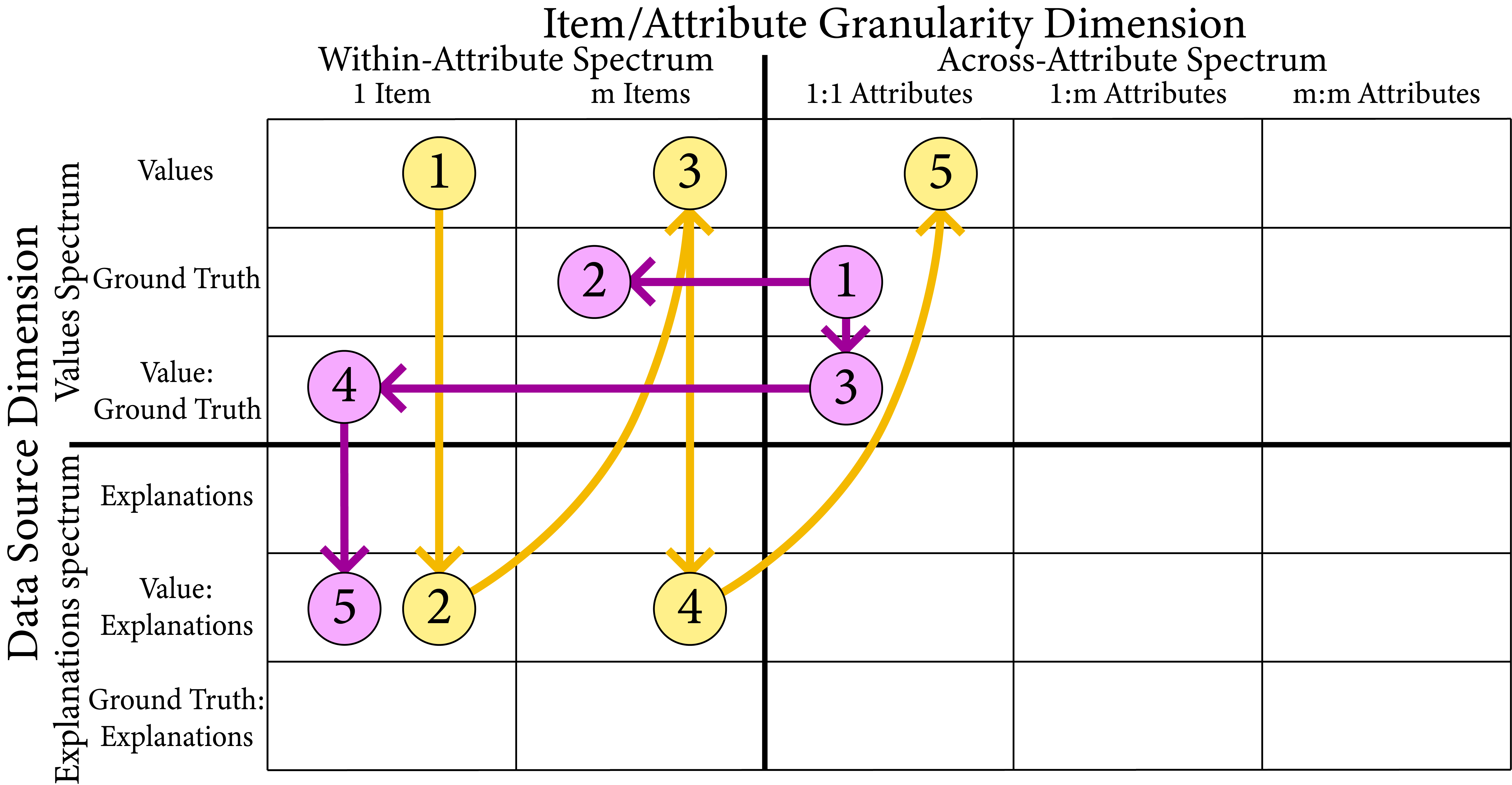}
\caption{
Mapping the supported tasks and user workflows of two existing validation approaches to our design space. 
Purple: iScore \cite{coscia2024iscore}, Yellow: LLM Comparator \cite{kahng2024llm}}
\vspace{-3mm}
\label{fig:validation_example}
\end{figure}

\begin{figure}[t]
\centering\includegraphics[width=1.0\linewidth]{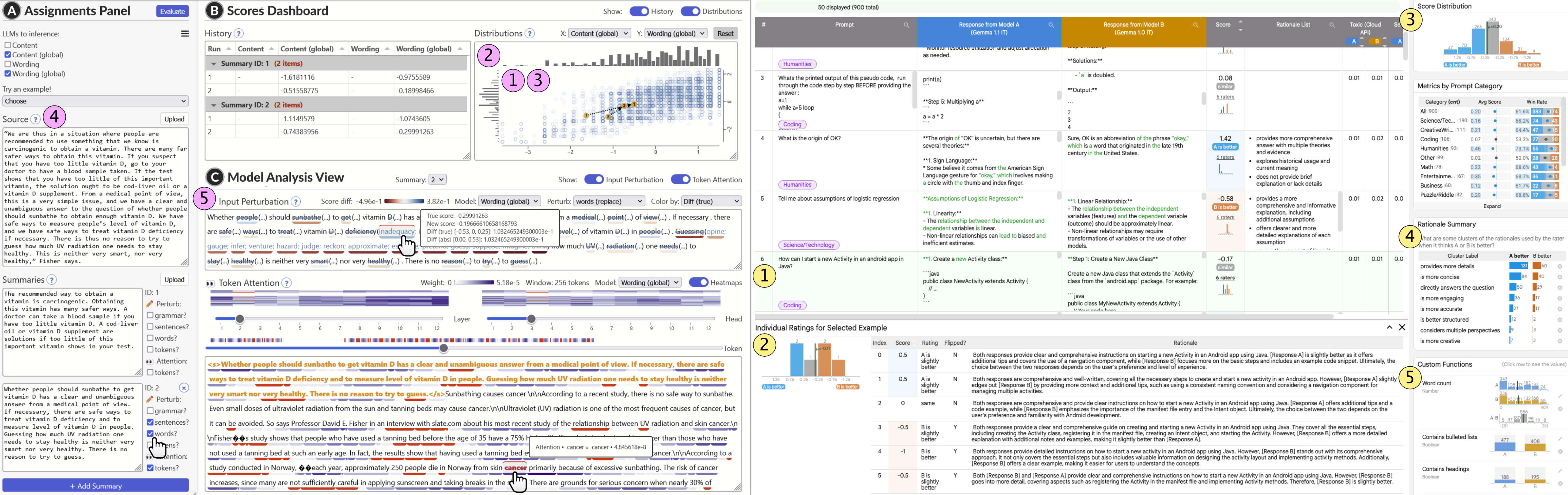}
\caption{
\changed{Screenshot with marked validation steps of iScore reproduced from Coscia et al.\cite{coscia2024iscore} with permission (left) and LLM Comparator \cite{kahng2024llm} \href{http://www.apache.org/licenses/LICENSE-2.0}{with permission} (right)}}
\vspace{-3mm}
\label{fig:validation_example}
\end{figure}

\textbf{iScore}~\cite{coscia2024iscore} is an LLM-based tool for scoring summaries on attributes, such as objectivity and coherence. 
A scatter plot displays ground truth data for two scoring attributes, providing an overview of expected trends and correlations \valExample{iScore}{(1)}.
A bar chart alongside the scatter plot further clarifies value distributions \valExample{iScore}{(2)}.
Users provide summaries which the LLM scores and visualizes, on the same scatter plot in a different color. 
Overlaying these plots allows direct comparison of trends between LLM scores and ground truth values \valExample{iScore}{(3)}.
To explore individual relationships, users can select a ground truth value, load the corresponding summary, and have it scored by the LLM for direct attribute comparison \valExample{iScore}{(4)}.
For deeper insight into how a score was derived, users can visualize the LLM’s attention and receive a non-textual explanation \valExample{iScore}{(5)}.

\textbf{LLM Comparator}~\cite{kahng2024llm} compares the output quality of different LLMs and identifies trends across models. 
Despite its primary focus on LLM-generated text, LLM Comparator uses 
an LLM-as-a-judge approach, based on pairwise comparisons, which generates a categorical gradation. 
Users select an item, view its prompt and two generated outputs, and examine individual comparison results \valExample{Comparator}{(1)}.
To understand why one output is preferred, textual explanations are collected during the LLM-as-a-judge process \valExample{Comparator}{(2)}.
For a broader perspective, users can analyze the distribution of aggregated scores to detect patterns in LLM preferences \valExample{Comparator}{(3)}.
Performance contextualization is possible through bar graphs linking explanations--such as ''is more engaging''--to output preferences \valExample{Comparator}{(4)}.
In the final step, users explore how LLM preferences relate to collected output metrics, such as word count, moving beyond within-attribute analysis to uncover \changed{relations} and trends \valExample{Comparator}{(5)}.
This structured comparison allows users to assess LLM performance, understand decision rationales, and detect potential biases or inconsistencies in model outputs.

\section{Limitations and Discussion}

\hspace{\parindent}\textbf{Data Source Mapping}: 
When validation approaches use two Data Source Dimension elements (values, ground truth, explanations), we currently map them only to the combined cell rather than separately, leading to empty explanation-only cells never occurring in isolation.
This reduces redundancy, but affects the descriptive nature of our mapping strategy, which we plan to investigate further.

\textbf{Item/Attribute Granularity Mapping:} 
Some validation approaches support both fine-grained and coarse analysis in one view. 
For example, scatter plots enable per-item, multi-item, and 1:1 attribute validation~\cite{coscia2024iscore}. 
Our mapping prioritizes the coarser cell to reduce redundancy, but this may impact the descriptiveness of the design space, requiring further investigation.


\textbf{Human Knowledge Externalization}: 
A key validation strategy compares LLM-generated data with human-provided ground truth, leveraging domain knowledge for representative subsets.
However, this approach faces scalability challenges, as human knowledge externalization is labor-intensive. 
Addressing these limitations will be crucial for making human-in-the-loop validation feasible at scale.

\textbf{Generation-Validation-Application as a Process}: 
We identify a structured process-oriented perspective for validating LLM-generated tabular data, with three key phases: data generation (including prompt engineering), validation (as studied), and downstream applications. 
Expanding the design space could distinguish validation for upstream generation and downstream application, forming a unified framework adaptable to specific domains.

\textbf{Numeric vs. Categorical Values}: While our design space is attribute-type agnostic, our illustrations focus on numerical representations of LLM-generated data values, \changed{and their relations~\cite{cagSachdeva2023}}. 
Future work may explore illustrative support for categorical relation discovery\changed{~\cite{cgf2014jb}} and develop idioms to an equal status, to ease the ideation of solutions for categorical data. 

\changed{\textbf{Role of Visual Analytics}:
Most pioneering validation approaches included VA building blocks to cope with the complexity of validation challenges.
This motivates a deeper study of forms of human involvement in combination with possibilities for computational support.
Future work includes the extension of the design space with a systematic focus on human-model collaboration patterns, studied in relation to the two existing dimensions.
}
\vspace{-4mm}
\section{Conclusion}
We introduced a structured design space for the critical validation of LLM-generated tabular data.
The design space systematically characterizes validation approaches, enabling their comparison and guiding developers in identifying validation strategies.
We mapped existing approaches onto the design space to identify their characteristics and discussed two approaches in detail to validate the descriptive power. \changed{Our work makes a case for VA as a critical layer in the LLM data generation. While automated solutions are valuable, interactive visual validation tools are indispensable for ensuring trust and reliability in validating synthetic tabular data. This design space provides a potential foundation for building such tools.}
Future work may explore extensions of the design space with further dimensions that could enrich validation approaches. 
The design space can also be expanded by formulating a unified process framework, distinguishing validation methods for data generation and application. 
\bibliographystyle{eg-alpha} 
\bibliography{01_main}       


\end{document}